\documentclass[a4paper,12pt]{article}

\usepackage{amssymb,amsmath}
\usepackage{mathrsfs}
\usepackage{bbm}

\allowdisplaybreaks \numberwithin{equation}{section}
\newcommand{\set}[1]{\left\{#1\right\}}
\newcommand{\pair}[1]{\left\langle#1\right\rangle}

\newtheorem{thm}{Theorem}[section]
\newtheorem{prp}[thm]{Proposition}
\newtheorem{lem}[thm]{Lemma}
\newtheorem{defn}[thm]{Definition}
\newenvironment{dfn}{\begin{defn} \rm }{\end{defn}}

\newtheorem{example}[thm]{Example}
\newenvironment{exa}{\begin{example} \rm }{\end{example}}
\newtheorem{remark}[thm]{Remark}

\newenvironment{rmk}{\begin{remark} \rm }{\end{remark}}
\newenvironment{prf}{{\it Proof.}}{\hfill$\Box$}

\newcommand\od{\mathrm{d}}

\newcommand{\nn}{\nonumber}

\newcommand{\bm}[1]{\hbox{\boldmath{$#1$}}}

\newcommand\ep{\epsilon}
\newcommand\pd{\partial}

\newcommand{\al}{\alpha}

\newcommand{\sg}{\sigma}
\newcommand{\om}{\omega} 

\newcommand{\Ld}{\Lambda}

\newcommand{\dt}{\delta}

\newcommand{\res}{\mathrm{res}} \newcommand{\Res}{\mathrm{Res}}

\newcommand\C{\mathbb{C}}

\newcommand\Z{\mathbb{Z}}
\newcommand\Zop{\mathbb{Z^{\mathrm{odd}}_+}}

\newcommand\cA{\mathcal{A}}
\newcommand\cD{\mathcal{D}}
\newcommand\cE{\mathcal{E}}

\newcommand\cG{\mathcal{G}}\newcommand\cH{\mathcal{H}}
\newcommand\cL{\mathcal{L}}
\newcommand\cM{\mathcal{M}}

\newcommand\cP{\mathcal{P}}
\newcommand\cU{\mathcal{U}}
\newcommand\cV{\mathcal{V}}

\newcommand\fD{\mathfrak{D}}
\newcommand\fE{\mathfrak{E}}
\newcommand\fg{\mathfrak{g}}

\newcommand\ra{\rangle}
\newcommand\la{\langle}

\begin{document}
\title{$R$-matrices and Hamiltonian Structures for \\ Certain Lax Equations}
\author{
Chao-Zhong Wu \thanks{Email address: wucz@sissa.it, Tel: +39 040
3787 352, Fax: +39 040 3787 466}
\\
 {\small Marie Curie fellow of the Istituto Nazionale di Alta
Matematica}
 \\
 {\small SISSA, via Bonomea 265, 34136 Trieste, Italy }
 }
\date{}
\maketitle

\begin{abstract}
In this paper a list of $R$-matrices on a certain coupled Lie
algebra is obtained. With one of these $R$-matrices, we construct
infinitely many bi-Hamiltonian structures for each of the
two-component BKP and the Toda lattice hierarchies. We also show
that, when such two hierarchies are reduced to their subhierarchies,
these bi-Hamiltonian structures are reduced correspondingly.

\end{abstract}


\section{Introduction}

The existence of Hamiltonian (or Poisson) structures reveals very
important property of a nonlinear evolutionary equation (see, for
example, \cite{Dickey}). For an evolutionary equation written in Lax
form, an efficient way to endow it with Hamiltonian structures is
the so-called classical $R$-matrix formalism. The classical
$R$-matrix formalism was proposed by Semenov-Tyan-Shanskii
\cite{STS} to construct Poisson brackets on a Lie algebra of an
associative algebra where the Lax equations are defined. Generally,
Semenov-Tyan-Shanskii's method yields two compatible Poisson
brackets, i.e., any linear combination of them is still a Poisson
bracket. Consequently one obtains a bi-Hamiltonian structure of the
Lax equation by performing a Dirac reduction \cite{MR} if needed.
Such a formalism was first established for anti-symmetric
$R$-matrices satisfying the modified Yang-Baxter equation \cite{RS}.
As later developed by Li and Parmentier \cite{LP}, also by Oevel and
Ragnisco \cite{OR}, the $R$-matrix formalism becomes available for a
wider class of $R$-matrices. Moreover, this method can produce three
compatible Poisson brackets on an associative algebra.

In this paper we study $R$-matrices on a ``coupled'' Lie algebra of
the form $\fg=\cG^-\times\cG^+$, whose Lie bracket is defined
diagonally by the Lie brackets on $\cG^\pm$ (see Section~3 below).
By solving the modified Yang-Baxter equation on $\fg$, we will
derive a class of $R$-matrices, which include the $R$-matrices used
in \cite{Carlet, BA, DSk, WX} as particular cases. Our goal is to
choose an appropriate $R$-matrix and apply the approach in \cite{LP,
OR} to construct Hamiltonian structures for Lax equations defined on
$\fg$. Two typical examples of such Lax equations, the two-component
BKP hierarchy \cite{DJKM-KPtype, LWZ} and the Toda lattice hierarchy
\cite{UT}, will be considered.

The two-component BKP hierarchy, i.e., the two-component
Kadomtsev-Petviashvili (KP) hierarchy of type B, was proposed by
Date, Jimbo, Kashiwara and Miwa \cite{DJKM-KPtype} as a bilinear
equation in consideration of a neutral free fermions realization of
the basic representation of the Lie algebra $D_\infty$. This
hierarchy was found to characterize Prym varieties in algebra
geometry \cite{Sh} and D-type topological Landau-Ginzburg models
\cite{Ta}; it is also known as the universal hierarchy of
Drinfeld-Sokolov hierarchies of type D, which are bi-Hamiltonian
equations associated to untwisted affine Kac-Moody algebra
$D_n^{(1)}$ with the zeroth vertex $c_0$ of its Dynkin diagram
marked \cite{DS, LWZ}. Recently we represented the two-component BKP
hierarchy into a Lax form \cite{LWZ} (cf. \cite{Sh}) with two types
of pseudo-differential operators. The sets $\cD^\pm$ of these two
types of operators compose a coupled Lie algebra $\cD^-\times\cD^+$
of the form $\fg$. Based on the Lax representation, in \cite{WX}  we
derived a bi-Hamiltonian structure for the two-component BKP
hierarchy by using the $R$-matrix formalism. However, from this
bi-Hamiltonian structure we could not find the corresponding
reductions when the two-component BKP hierarchy is reduced to
Drinfeld-Sokolov hierarchies mentioned above.  To resolve this
problem, we will choose an $R$-matrix that is different from the one
in \cite{WX} on $\fg$, then construct a series of bi-Hamiltonian
structures for the two-component BKP hierarchy. An advantage of
these bi-Hamiltonian structures is that, they can be reduced to the
bi-Hamiltonian structures for the corresponding Drinfeld-Sokolov
hierarchies of type D. Other reductions of the Hamiltonian
structures for the two-component BKP hierarchy will also be studied,
see Section~4 below.

As the second example, the Toda lattice hierarchy \cite{UT} has a
Lax representation on a coupled Lie algebra of the form $\fg$
consisting of shift operators. For this hierarchy, there are three
compatible Hamiltonian structures found by Carlet \cite{Carlet}, but
the reduction property of their Hamiltonian structures has not been
considered before. By using an $R$-matrix introduced in the present
paper, we will show that the Toda lattice hierarchy possesses
infinitely many bi-Hamiltonian structures labeled by arbitrary
positive-integer pairs $(N,M)$. Particularly, such Hamiltonian
structures in the case $N=M=1$ are similar, but not the same, with
those given in \cite{Carlet}. Furthermore, we will show that the
bi-Hamiltonian structure labeled by $(N,M)$ for the Toda lattice
hierarchy is reduced to that for the extended $(N,M)$-bigraded Toda
hierarchy \cite{Ca06, CDZ} under suitable constraint.

The bi-Hamiltonian structures to be obtained for the two-component
BKP and the Toda lattice hierarchies, in comparison with those in
\cite{Carlet, WX}, have more advantages. First, the densities of
Hamiltonian functionals satisfy the tau-symmetry condition
\cite{DZ}, hence they define a tau function of the hierarchy.
Second, these bi-Hamiltonian structures have important application
in the study of Frobenius manifolds. In the finite-dimensional case,
Frobenius manifold was proposed by Dubrovin \cite{Du} as a
coordinate-free description of the WDVV equation in 2\,D topological
field theory. Generally speaking, associated to every Frobenius
manifold there is a bi-Hamiltonian structure of hydrodynamic type,
which links integrable hierarchies with relevant research branches
of mathematical physics. The theory of Frobenius manifold was
extended to the infinite-dimensional case by Carlet, Dubrovin and
Mertens \cite{CDM} in consideration of a bi-Hamiltonian structure
for the dispersionless Toda lattice hierarchy, which coincides with
the dispersionless limit of the bi-Hamiltonian structure with
$N=M=1$ in Section~\ref{sec-TL} below. Following the approach of
\cite{CDM}, we constructed a class of infinite-dimensional Frobenius
manifolds underlying the bi-Hamiltonian structures obtained in the
paper for the dispersionless two-component BKP hierarchy \cite{WX2}.
Furthermore, these infinite-dimensional Frobenius manifolds contain
finite-dimensional Frobenius submanifolds corresponding to the
reductions of bi-Hamiltonian structures.

This paper is arranged as follows. In next section we review the
$R$-matrix formalism for the construction of Poisson brackets. In
Section~3 we derive a list of $R$-matrices on the Lie algebra
$\fg=\cG^-\times\cG^+$, which are classified according to the action
of some simple intertwining involutions. With one of these
$R$-matrices, we apply the $R$-matrix formalism to the two-component
BKP and the Toda lattice hierarchies in Section~4 and Section~5
respectively. The reduction property of these Hamiltonian structures
will be clarified. The final section is devoted to the conclusion.

\section{Classical $R$-matrices}

Let us recall some properties of classical $R$-matrices, and how to
employ them to construct Poisson bracket on a Lie algebra.

\subsection{The $R$-matrix formalism}

All contents in this subsection can be found in \cite{Li, LP, OR,
STS}.

Let $\fg$ be a complex Lie algebra.
\begin{dfn}
A linear transformation $R: \fg\to\fg$ is called an $R$-matrix if it
defines a Lie bracket as
\begin{equation}\label{}
[X,Y]_R=[R(X), Y]+[X, R(Y)],\quad X, Y\in\fg.
\end{equation}
\end{dfn}
A sufficient condition for a linear transformation $R$ being an
$R$-matrix is that it solves the following modified Yang-Baxter
equation
\begin{equation}\label{YBeq}
[R(X), R(Y)]-R([X, Y]_R)=-[X,Y]
\end{equation}
for any $X$, $Y\in\fg$.

Assume $\fg$ to be an associative algebra, on which a Lie bracket is
defined naturally by the commutator. We also assume that there is a
function $\la~\ra:\fg\to\C$, and it defines a non-degenerate
symmetric invariant bilinear form (inner product) $\la~ ,\,\ra$ by
\[
\langle X, Y\rangle=\la X Y\ra=\la Y X\ra,\quad X, Y\in\fg.
\]
Via this inner product $\fg$  can be identified with its dual space
$\fg^*$. Let $T\fg$ and $T^*\fg$ denote the tangent and the
cotangent bundles of $\fg$, whose fibers at any point $A$ are
$T_A\fg=\fg$ and $T^*_A\fg=\fg^*$  respectively.

Given an $R$-matrix $R$ on $\fg$, let $R^*$ be the adjoint
transformation of $R$ with respect to the above inner product. The
anti-symmetric part of $R$ reads
\[
R_a=\frac1{2}(R-R^*).
\]
For any smooth functions $f, g\in C^\infty(\fg)$, there are three
brackets:
\begin{align} \label{linbr}
\{f, g\}_1(A)&=\frac1{2}\big(\la[A,\od f],R(\od
g)\ra-\la[A,\od g],R(\od f)\ra\big), \\
 \{f, g\}_2(A)&=\frac1{4}\big(\la[A,\od f],R(A\cdot\od
g+\od g\cdot A)\ra-\la[A,\od g],R(A\cdot\od f+\od f\cdot
A)\ra\big), \label{quabr} \\
\{f, g\}_3(A)&=\frac1{2}\big(\la[A,\od f],R(A\cdot\od g\cdot
A)\ra-\la[A,\od g],R(A\cdot\od f\cdot A)\ra\big), \label{cubbr}
\end{align}
where $\od f, \od g \in T_A^*\fg$ are the gradients of $f, g$ at
$A\in\fg$ respectively. The brackets \eqref{linbr}--\eqref{cubbr}
are called the linear, the quadratic and the cubic brackets
respectively.

\begin{thm}[\cite{LP, OR}]\label{thm-3br}
The following statements hold true:
\begin{itemize}
  \item[(1)] for any $R$-matrix $R$, the linear bracket is a Poisson
bracket;
\item[(2)] if both $R$ and its anti-symmetric part
$R_a$ solve the modified Yang-Baxter equation \eqref{YBeq}, then the
quadratic bracket is a Poisson bracket;
\item[(3)]
if $R$ satisfies the modified Yang-Baxter equation \eqref{YBeq} then
the cubic bracket is a Poisson bracket.
\end{itemize}
Moreover, these three Poisson brackets are compatible whenever all
the above conditions are fulfilled.
\end{thm}

In case the associative algebra $\fg$ is non-commutative, then the
$R$-matrix formalism gives no more Poisson brackets of order higher
than $3$. However, if $\fg$ is a commutative associative algebra,
then one can have Poisson brackets with order being any positive
integers due to the following theorem.
\begin{thm}[\cite{Li}] \label{thm-Li}
Let $\fg$ be a Poisson algebra, that is, a Lie algebra of a
commutative associative algebra with Lie bracket satisfying
$[X,Y\,Z]=[X,Y]Z+Y[X,Z]$ for all $X, Y, Z\in\fg$, and endow $\fg$
with an ad-invariant inner product $\la~ ,\,\ra$ being invariant
with respect to the multiplication, i.e., $\la X Y,Z\ra=\la X, Y
Z\ra$. If $R$ is an $R$-matrix on $\fg$, then there exists a series
of compatible Poisson brackets defined as follows:
\begin{equation}\label{pbr}
\{f,g\}_r(A)=\frac1{2}\left(\la[A,\od f],R(A^{r-1}\od
g)\ra-\la[A,\od g],R(A^{r-1}\od f)\ra\right)
\end{equation}
with $f,g\in C^\infty(\fg)$ and arbitrary positive integers $r$.
\end{thm}

\subsection{Intertwining operators}

A linear operator $\sg: \fg\to\fg$ is said to be intertwining if it
satisfies
\begin{equation}\label{sg}
\sg[X,Y]=[\sg X,Y]=[X,\sg Y], \quad X,Y\in\fg.
\end{equation}
\begin{prp}[\cite{RS}]
If $R$ is an $R$-matrix and $\sg$ is an intertwining operator, then
$R\circ\sg$ is also an $R$-matrix.
\end{prp}
Note that all intertwining operators compose a linear family. Hence
the $R$-matrices $R\circ\sg$, with $R$ fixed and $\sg$ being
intertwining operators, induce a family of compatible Poisson
brackets.

\begin{dfn}
A linear operator $\sg: \fg\to\fg$ is called an intertwining
involution if it satisfies \eqref{sg} and $\sg\circ\sg=\mathrm{id}$.
\end{dfn}

\begin{prp}\label{thm-intertw}
Let $R$ be a solution of the modified Yang-Baxter equation
\eqref{YBeq} and $\sg$ be an intertwining involution, then
$R\circ\sg$ also solves equation \eqref{YBeq}.
\end{prp}
\begin{prf}
This proposition follows from a simple calculation:
\begin{align*}\label{}
&[R\circ\sg X, R\circ\sg Y]-R\circ\sg([R\circ\sg X, Y]+[X,R\circ\sg
Y]) \\
=& [R\circ\sg X, R\circ\sg Y]-R([R\circ\sg X, \sg Y]+[\sg X,
R\circ\sg Y]) \\
=&-[\sg X, \sg Y]= -\sg\circ\sg [X,Y] =-[X,Y].
\end{align*}
\end{prf}

\section{$R$-matrices on a coupled Lie algebra}\label{sec-R}

Let $\cG$ be a complex linear space that contains two subspaces
$\cG^-$ and $\cG^+$. On each $\cG^\pm$ there is a Lie bracket, and
these two Lie brackets coincide with each other when restricted on
$\cG^-\cap\cG^+$. Moreover, we assume that $\cG^\pm$ admit the
following decompositions of Lie subalgebras:
\begin{equation}\label{decom}
\cG^-=(\cG^-)_-\oplus(\cG^-)_+, \quad
\cG^+=(\cG^+)_-\oplus(\cG^+)_+,
\end{equation}
and
\[
(\cG^-)_+\subset(\cG^+)_+, \quad (\cG^+)_-\subset(\cG^-)_-.
\]
Introduce a coupled Lie algebra
\begin{equation}\label{g}
\fg=\cG^-\times\cG^+
\end{equation}
whose Lie bracket is defined diagonally by the brackets on $\cG^\pm$
as
\[
[(X,\hat{X}),\, (Y,\hat{Y})]=([X,Y],\,[\hat{X},\hat{Y}]), \quad
(X,\hat{X}),\, (Y,\hat{Y})\in\fg.
\]

Consider linear transformations $R: \fg\to\fg$ of the form
\begin{equation}\label{Rmat}
R(X,\hat{X})=(a\,X_+ +b\,X_- +c\,\hat{X}_-,d\,\hat{X}_+
+e\,\hat{X}_- +f\, X_+)
\end{equation}
with $a, b, c, d, e, f\in\C$. Here and below we use the subscripts
``$\pm$'' to denote the projections onto the Lie subalgebras
$(\cG^-)_\pm$ or $(\cG^+)_\pm$ respectively.
\begin{prp}\label{thm-rmat}
The transformation $R$ in \eqref{Rmat} solves the modified
Yang-Baxter equation \eqref{YBeq} if and only if $(a, b, c, d, e,
f)$ is one of the following:
\begin{align}
&  \pm(1,-1,-2,1,-1,2), \quad \pm(1,-1,2,-1,1,2), \label{case1}\\
&\pm(1,-1,0,1,1,2), \quad \pm(1,-1,0,-1,-1,2), \\
&\pm(1,1,-2,1,-1,0), \quad \pm(1,1,2,-1,1,0), \label{case3} \\
&(\pm1,\pm1,0,\pm1,\pm1,0). \label{case4}
\end{align}
Each of these solutions gives an $R$-matrix on the Lie algebra
$\fg$.
\end{prp}
\begin{prf}
For any $\bm{X}=(X,\hat{X}), \bm{Y}=(Y,\hat{Y})\in\fg$ and $R$ in
\eqref{Rmat}, the modified Yang-Baxter equation
\begin{equation*}
[R(\bm{X}),R(\bm{Y})]-R([\bm{X},\bm{Y}]_R)=-[\bm{X},\bm{Y}]
\end{equation*}
is expanded to
\begin{align*}
-(&a^2[X_+,Y_+]+a^2[X_+,Y_-]+a^2[X_-,Y_+]+b^2[X_-,Y_-]
\\ &-(a^2-b^2)([X_+,Y_-]_- +[X_-,Y_+]_-)
+c(-a+b+f)([X_+,\hat{Y}_-]_- +[\hat{X}_-,Y_+]_-)\\
&-c(d+e)[\hat{X},\hat{Y}]_-+c(e-d-c)[\hat{X}_-,\hat{Y}_-], \\
&d^2[\hat{X}_+,\hat{Y}_+]+d^2[\hat{X}_+,\hat{Y}_-]+d^2[\hat{X}_-,\hat{Y}_+]+e^2[\hat{X}_-,\hat{Y}_-]\\
&-(d^2-e^2)([\hat{X}_+,\hat{Y}_-]_-+[\hat{X}_-,\hat{Y}_+]_-)
+f(d-e+c)([X_+,\hat{Y}_-]_++[\hat{X}_-,Y_+]_+) \\
&+f(a+b)[X,Y]_++f(a-b-f)[X_+,Y_+])
\\
=-(&[X_+ +X_-,Y_+ +Y_-], [\hat{X}_+ +\hat{X}_-,\hat{Y}_+
+\hat{Y}_-]).
\end{align*}
Comparing the coefficients on both sides, we have
\begin{align}\label{}
&a^2=b^2=d^2=e^2=1, \\
&c(-a+b+f)=0, \quad c(e-d-c)=0, \quad c(d+e)=0, \\
&f(a-b-f)=0, \quad f(d-e+c)=0, \quad f(a+b)=0.
\end{align}
All solutions of these equations are listed in
\eqref{case1}--\eqref{case4}. Thus the proposition is proved.
\end{prf}

There are two intertwining involutions $\sg_1$ and $\sg_2$ on $\fg$
defined naturally by
\begin{equation}\label{sg12}
\sg_1(X,\hat{X})=(-X,\hat{X}), \quad \sg_2(X,\hat{X})=(X,-\hat{X}).
\end{equation}
They generate a group $G=\set{\mathrm{id}, \sg_1, \sg_2,
\sg_1\circ\sg_2}$ of intertwining involutions. Up to the action of
$G$ (see Proposition~\ref{thm-intertw}), the solutions in each line
of \eqref{case1}--\eqref{case3} are equivalent, while the solutions
in line \eqref{case4} are divided to four equivalence classes.

Among the $R$-matrices given in Proposition~\ref{thm-rmat}, from now
on we fix an $R$ as
\begin{equation}\label{Rmatrix}
R(X,\hat{X})=(X_+ -X_- -2\hat{X}_-,\hat{X}_+ -\hat{X}_- +2 X_+).
\end{equation}
This is the $R$-matrix that will be applied to construct Hamiltonian
structures for integrable hierarchies whose Lax representation is
defined on a coupled Lie algebra of the form \eqref{g}. Before
carrying out the construction, let us give some remarks at the end
of this section.

\begin{rmk}\label{rmk-tR}
The $R$-matrix corresponding to the solution $(1,-1,2,-1,1,2)$ in
\eqref{case1}, denoted as $\tilde{R}$, was introduced by Carlet
\cite{Carlet} on a coupled Lie algebra of shift operators (see
Section~5 below). This is the $R$-matrix used in \cite{Carlet, WX}
to construct Hamiltonian structures for the Toda lattice and the
two-component BKP hierarchies. Observe that $\tilde{R}=R\circ\sg_2$,
where $R$ is given in \eqref{Rmatrix}.
\end{rmk}

\begin{rmk}\label{}
Every solution in line \eqref{case4} splits into $R$-matrices on
$\cG^-$ or $\cG^+$; as a corollary of Proposition~\ref{thm-rmat},
they are the only solutions of the form $a X_+ +b X_-$ or $d
\hat{X}_+ +e \hat{X}_-$ to the modified Yang-Baxeter equation.
\end{rmk}

\begin{rmk}\label{}
When $\cG^-=\cG^+$, an $R$-matrix of the form \eqref{Rmatrix} was
used in \cite{BA} to prove the Liouville integrability of the Toda
lattice defined on semi-simple Lie algebras. In this case,
$\cG^-\times\cG^-$ is called the classical double of the Lie algebra
$\cG^-$. On such a classical double, Dubrovin and Skrypnyk
\cite{DSk} studied the ``Adler-Kostant-Symes''
$R$-operators/matrices of the form $\tilde{R}$ and the corresponding
commutative hamiltonian flows. It would be interesting to consider
the action of intertwining involutions on such $R$-operators.
\end{rmk}

\section{Hamiltonian structures for the two-component BKP hierarchy}

Let us employ the $R$-matrix \eqref{Rmatrix} to construct
Hamiltonian structures for the two-component BKP hierarchy. What is
more, the reduction property of these Hamiltonian structures will be
studied.

\subsection{Pseudo-differential operators and Lax representation}

We first review the definition of the two-component BKP hierarchy
and some necessary notations.

Let $\cA$ be an algebra of smooth functions of $x\in S^1$, on which
there is a derivation $D=\od/\od x$. Assume $\mathcal{A}$ to be
graded as $\mathcal{A}=\prod_{i\geq0}\cA_i$ such that
$\cA_i\cdot\cA_j\subset\cA_{i+j}$ and $D(\cA_i)\subset\cA_{i+1}$.
Denote $\cD=\set{\sum_{i\in\Z} f_i D^i\mid f_i\in\cA}$ and consider
its two subspaces
\begin{align}
&\cD^-=\left\{ \sum_{i<\infty} f_i D^i \mid f_i\in\cA\right\}, \\
&\cD^+=\left\{ \sum_{i\in\Z}\sum_{j\ge \max\{0,m-i\}}a_{i,j} D^i
\mid a_{i,j}\in\cA_j, m\in\Z \right\}.
\end{align}
The subspaces $\cD^-$ and $\cD^+$, equipped with a product defined
by
\begin{equation}\label{pro}
f D^i\cdot g D^j=\sum_{r\geq0}\binom{i}{r}f\, D^r(g)\, D^{i+j-r},
\quad f,\,g\in\cA,
\end{equation}
are called the algebras of pseudo-differential operators of the
first type and the second type respectively \cite{LWZ}. Observe that
every operator in $\cD^-$, of no difference from a usual
pseudo-differential operator \cite{Dickey}, has an upper bound for
the powers in $D$. For an operator in $\cD^+$, there may be neither
an upper nor a lower bound for the powers in $D$, but as the power
in $D$ decreases the degree of the coefficient must increase
simultaneously, which makes $\cD^+$ be closed for the product
\eqref{pro}.

Given a pseudo-differential operator $A=\sum_{i\in\Z} f_i
D^i\in\cD^\pm$, its positive part, negative part, residue and
adjoint operator are defined respectively by
\begin{align}\label{Apm}
&A_+=\sum_{i\geq0} f_i  D^i, \quad A_-=\sum_{i<0} f_i  D^i, \\
&\res\,A=f_{-1},\quad A^*=\sum_{i\in\Z}(- D)^i\cdot f_i.
\end{align}
The projections given in \eqref{Apm} induce the following
decompositions of  subalgebras:
\begin{equation}\label{Dpm}
\cD^\pm=(\cD^\pm)_+\oplus(\cD^\pm)_-.
\end{equation}
Clearly $(\cD^-)_+\subset(\cD^+)_+$ and $(\cD^+)_-\subset(\cD^-)_-$.

Assume $\{u_1,u_3,u_5,\dots,
\hat{u}_{-1},\hat{u}_1,\hat{u}_3,\dots\}$ to be a set of independent
functions in $\cA_0\subset\cA$; particularly, we assume
$\hat{u}_{-1}\ne0$. Introduce two pseudo-differential operators
\begin{equation} \label{PhP}
P= D+\sum_{i\ge1}u_i  D^{-i}, \quad \hat{P}=
D^{-1}\hat{u}_{-1}+\sum_{i\ge1}\hat{u}_i D^i
\end{equation}
 such that
\[
 P^*=- D P D^{-1}, \quad \hat{P}^*=- D\hat{P}
D^{-1}.
\]
Note that $P\in\cD^-$, and $\hat{P}\in\cD^+$ for
$D^{-1}\hat{u}_{-1}=\sum_{i\ge0}(-D)^i(\hat{u}_{-1})D^{-i-1}$.

\begin{dfn}
The two-component BKP hierarchy is defined by the following Lax
equations \cite{LWZ}:
\begin{align}\label{PPht}
& \frac{\pd P}{\pd t_k}=[(P^k)_+, P], \quad \frac{\pd \hat{P}}{\pd t_k}=[(P^k)_+, \hat{P}],  \\
\label{PPhth} & \frac{\pd P}{\pd \hat{t}_k}=[-(\hat{P}^k)_-, P],
\quad \frac{\pd \hat{P}}{\pd \hat{t}_k}=[-(\hat{P}^k)_-, \hat{P}]
\end{align}
with $k\in\Zop$.
\end{dfn}
The name of this hierarchy is from its equivalent version of
bilinear equation constructed by Date, Jimbo, Kashiwara and Miwa
\cite{DJKM-KPtype}, see equation~\eqref{bltau} below.

To study Hamiltonian structures for the two-component BKP hierarchy
, we need further preparation.

By a local functional on $\cA$ we mean an element of the quotient
space $\cA/ D(\cA)$, written formally as $\int f\od x$ with
$f\in\cA$. Introduce a map
\begin{equation}\label{int}
\la\, \,\ra:\ \cD\to\cA/ D(\cA),\quad A\mapsto \la A\ra=\int \res
A\, \od x.
\end{equation}
It can be checked that this map induces an inner product on each of
$\cD^\pm$ by
\begin{equation} \label{inpro}
\la A,B\ra=\la A B\ra=\la B A\ra.
\end{equation}
For any subspace $\mathcal{S}\subset\cD^\pm$, let $\mathcal{S}^*$
denote its dual space with respect to the inner product
\eqref{inpro}. For example, one has
\begin{equation}\label{Ddual}
(\cD^\pm)^*=\cD^\pm, \quad \big((\cD^\pm)_\pm\big)^*=(\cD^\pm)_\mp.
\end{equation}

The spaces $\cD^\pm$ can be decomposed as
\begin{equation}\label{Deo}
\cD^\pm=\cD^\pm_0\oplus\cD^\pm_1, \quad \cD^\pm_\nu=\left\{
A\in\cD^\pm\mid A^*=(-1)^\nu A\right\}.
\end{equation}
Note that the dual spaces of $\cD^\pm_\nu$ are
$(\cD^\pm_\nu)^*=\cD^\pm_{1-\nu}$. Every element of $\cD^\pm_\nu$
can be expressed in the form
\[
\sum_{i\in\Z}\left(a_i\,D^{2 i+\nu}+D^{2 i+\nu}a_i\right), \quad
a_i\in\cA,
\]
then for any $l\in\Z$ we have the following decompositions of
subspaces:
\begin{equation}\label{D01}
\cD^\pm_\nu=(\cD^\pm_\nu)_{\ge l}\oplus (\cD^\pm_\nu)_{<l}, \quad
\nu=0,1,
\end{equation}
where
\begin{align*}\label{}
&(\cD^\pm_\nu)_{\ge l}=\left\{\sum_{2 i+\nu\ge l}\left(a_i\,D^{2
i+\nu}+D^{2 i+\nu}a_i\right)\in \cD^\pm\mid a_i\in\cA\right\},
\\
& (\cD^\pm_\nu)_{< l}=\left\{\sum_{2 i+\nu< l}\left(a_i\,D^{2
i+\nu}+D^{2 i+\nu}a_i\right)\in \cD^\pm\mid a_i\in\cA\right\}.
\end{align*}

\subsection{Bi-Hamiltonian representations}

The two-component BKP hierarchy \eqref{PPht}--\eqref{PPhth} was
first represented into a bi-Hamiltonian form by us in \cite{WX}.
With the same method but the $R$-matrix $\tilde{R}$ (see
Remark~\ref{rmk-tR}) replaced by $R$ in \eqref{Rmatrix}, we want to
derive different bi-Hamiltonian structures for this hierarchy.

Recall the sets $\cD^\pm$ of pseudo-differential operators of the
first and the second type over $\cA$, and realize the coupled Lie
algebra $\fg$ in \eqref{g} as
\begin{equation}
\fD=\cD^-\times\cD^+.
\end{equation}
On $\fD$ the product and Lie bracket are defined diagonally, and
there is an inner product given by
\begin{equation}\label{inpD}
\la(X,\hat{X}),(Y,\hat{Y})\ra=\la(X,\hat{X})(Y,\hat{Y})\ra=\la X,
Y\ra+\la\hat{X},\hat{Y}\ra
\end{equation}
for any $(X,\hat{X}),(Y,\hat{Y})\in\fD$.

On the algebra $\fD$ we have the $R$-matrix $R$ given in
\eqref{Rmatrix}, in which the subscripts ``$\pm$'' mean the
projections \eqref{Dpm}. Since
\begin{align*}
\la R(X,\hat{X}),(Y,\hat{Y})\ra=&\la(X_+ - X_- -
\hat{X}_-)Y+(\hat{X}_+ - \hat{X}_- + X_+)\hat{Y}\ra \\
=&\la X(Y_- - Y_+ +\hat{Y}_-)+\hat{X}(-Y_+ + \hat{Y}_- -
\hat{Y}_+)\ra \\
=&\la (X,\hat{X}),-R(Y,\hat{Y})\ra,
\end{align*}
then $R^*=-R$, namely,  $R$ is anti-symmetric. Both $R$ and its
anti-symmetric part $R_a=R$ satisfy the modified Yang-Baxter
equation \eqref{YBeq}, hence Theorem~\ref{thm-rmat} can be applied
to endow $\fD$ with Poisson brackets.

Consider $\fD$ as an infinite-dimensional manifold, whose
coordinates are given by the coefficients of its elements
\begin{equation}\label{eqA}
\bm{A}=\left(\sum_{i\in\Z}w_i D^i,\, \sum_{i\in\Z}\hat{w}_i
D^i\right).
\end{equation}
For any local functional $F=\int f\,\od x$ on  $\fD$, its
variational gradient $\dt F/\dt\bm{A}$ at $\bm{A}$ is defined by
$\dt F=\la \dt F/\dt\bm{A},\dt\bm{A}\ra$. More explicitly,
\[
\frac{\dt F}{\dt\bm{A}}=\left(\sum_{i\in\Z} D^{-i-1}\frac{\dt F}{\dt
w_i},\, \sum_{i\in\Z}D^{-i-1}\frac{\dt F}{\dt\hat{w}_i}\right),
\]
where $\dt F/\dt w=\sum_{j\ge0}(-D)^j\left(\pd f/\pd
w^{(j)}\right)$. We emphasize that in this section only functionals
with their gradients lying in $\fD$ will be considered.

According to the second assertion in Theorem~\ref{thm-3br} we have
the following result.
\begin{lem}\label{}
Let $F$ and $H$ be two arbitrary functionals. On the algebra $\fD$
there is a quadratic Poisson bracket
\begin{equation}\label{Poibra}
\{F, H\}(\bm{A})=\left\la\frac{\dt F}{\dt\bm{A}},
\cP_{\bm{A}}\left(\frac{\dt H}{\dt\bm{A}}\right)\right\ra, \quad
\bm{A}=(A,\hat{A})\in\fD,
\end{equation}
 where the Poisson
tensor $\cP: T\fD^*\to T\fD$ is defined by
\begin{align}
\cP_{(A,\hat{A})}(X,\hat{X})=& \big(-(A X+\hat{A}\hat{X})_-A+A(X
A+\hat{X}\hat{A})_-, \nn\\
&\quad (A X+\hat{A}\hat{X})_+\hat{A}-\hat{A}(X
A+\hat{X}\hat{A})_+\big). \label{PAX}
\end{align}
\end{lem}
\begin{prf}
The bracket~\eqref{Poibra} is a reformulation of \eqref{quabr} with
$\fg=\fD$ and $R$ given in \eqref{Rmatrix}.
\end{prf}

In order to obtain Hamiltonian structures for the two-component BKP
hierarchy, we need to reduce the Poisson bracket \eqref{Poibra} to
appropriate submanifolds of $\fD$ where the flows
\eqref{PPht}--\eqref{PPhth} are defined.

First, according to \eqref{Deo}, one decomposes the space $\fD$ as
\begin{equation}\label{bddec}
\fD=\fD_0\oplus\fD_1, \quad \fD_\nu=\cD^-_\nu\times\cD^+_{\nu}
\hbox{ for } \nu=0,1.
\end{equation}
The subspaces $\fD_0$ and $\fD_1$ are dual to each other, hence at
any point $\bm{A}\in\fD_\nu$ one can identify
$T_{\bm{A}}^*\fD_\nu=(\fD_\nu)^*=\fD_{1-\nu}$. It is straightforward
to check the following lemma.
\begin{lem}\label{thm-qbrD}
 The  subspaces $\fD_0$ and $\fD_1$ are Poisson submanifolds of $\fD$
 with respect to the Poisson bracket \eqref{Poibra}.
\end{lem}

Second, for the operators \eqref{PhP} and an arbitrary positive
integer $m$, let
\begin{equation}\label{eqbK}
\bm{A}=(A,\hat{A})=(D P^{2 m}, D\hat{P}^2)
\end{equation}
with $P$ and $\hat{P}$ given in \eqref{PhP}. Clearly, one has
\begin{align}
&A=D^{2 m+1}+\sum_{i\le m}(v_{i} D^{2i-1}+f_{i} D^{2i-2}), \label{A}
\\
&\hat{A}=\rho D^{-1}\rho+\sum_{i\ge1}(\hat{v}_{i}
D^{2i-1}+\hat{f}_{i} D^{2i-2}), \quad \rho=\hat{u}_{-1}. \label{Ah}
\end{align}
Denote $\hat{v}_{0}=\rho^2$ and $\bm{v}=(v_{m}, v_{m-1}, \dots,
\hat{v}_0, \hat{v}_1,\dots)$. Since
\[
(A^*,\hat{A}^*)=(-(P^*)^{2m} D,-(\hat{P}^*)^2 D)=(-D P^{2 m},
-D\hat{P}^2)= -(A,\hat{A}),
\]
then the coefficients $f_{-i}$ and $\hat{f}_i$ are linear functions
of derivatives of $\bm{v}$. Conversely, given two operators $A$ and
$\hat{A}$ as \eqref{A}--\eqref{Ah} constrained by
$(A^*,\hat{A}^*)=-(A,\hat{A})$, according to \cite{Dickey, LWZ}
there exist uniquely operators $P=(D^{-1}A)^{1/2 m}\in\cD^-$ and
$\hat{P}=(D^{-1}\hat{A})^{1/2}\in\cD^+$ of the form \eqref{PhP}.
Hence the coordinates $\bm{v}$ and $\bm{u}=(u_1, u_3, \dots,
\hat{u}_{-1}, \hat{u}_1, \hat{u}_3, \dots)$ given in \eqref{PhP} can
be represented by each other; this transformation of coordinates is
called a Miura-type transformation. It means that the Lax equations
\eqref{PPht}--\eqref{PPhth} can be defined equivalently as
\begin{align}\label{At}
&\frac{\pd\bm{A}}{\pd t_k}=D\left[\left((P^k)_+,(P^k)_+\right),(P^{2
m}, \hat{P}^2)\right],
\\
&\frac{\pd\bm{A}}{\pd\hat{t}_k}=D\left[\left(-(\hat{P}^k)_-,-(\hat{P}^k)_-\right),(P^{2
m}, \hat{P}^2)\right]. \label{Ath}
\end{align}
They are Lax equations on the coset consisting of operators of the
form \eqref{eqbK}, that is,
\begin{equation}\label{eqU}
\cU_m=(D^{2\,m+1},0)+(\cD^-_1)_{< 2
m}\times\big((\cD^+_1)_{\ge0}\times \cM \big)
\end{equation}
(recall \eqref{D01}) with
\[
\cM=\{\rho D^{-1}\rho\mid\rho\in\cA, \rho\ne0\}.
\]
Here $\cM$ is considered as a $1$-dimensional manifold with
coordinate $\rho$, and its tangent spaces are
\begin{equation}
T_\rho \cM=\{\rho D^{-1}f+f D^{-1}\rho\mid f\in\cA\}.
\end{equation}
The tangent bundle of the coset $\cU_m$, denoted as $T\cU_m$, has
fibers
\begin{equation}\label{tanU}
T_{\bm{A}}\cU_m=(\cD^-_1)_{< 2 m}\times\big((\cD^+_1)_{\ge0}\oplus
T_\rho \cM \big).
\end{equation}
Their dual spaces
\begin{equation}\label{ctanU}
T_{\bm{A}}^*\cU_m=(\cD^-_0)_{\ge-2
m}\times\big((\cD^+_0)_{<-1}\oplus T_\rho^*\cM \big), \quad
T^*_\rho\cM=\cA
\end{equation}
compose the cotangent bundle $T^*\cU_m$ of $\cU_m$. One sees that a
functional $F$ on the coset $\cU_m$ has variational gradient in
$T_{\bm{A}}^*\cU_m$ as
\begin{equation*}\label{vgF}
\frac{\dt F}{\dt\bm{A}}=\frac1{2}\left(\sum_{i\le m}\left(\frac{\dt
F}{\dt v_{i}} D^{-2i}+ D^{-2i}\frac{\dt F}{\dt v_{i}}\right),
\sum_{i\ge0}\left(\frac{\dt F}{\dt \hat{v}_{i}} D^{-2i}+
D^{-2i}\frac{\dt F}{\dt \hat{v}_{i}}\right)\right).
\end{equation*}

\begin{lem}\label{lem-Poicoset}
On the coset
 $\cU_m$ that consists of operators of the form
 \eqref{eqbK}, the map $\cP: T^*\cU_m\to T\cU_m$ defined by \eqref{PAX}
is a Poisson tensor.
\end{lem}
\begin{prf}
We perform a Dirac reduction for $\cP$ from $\fD_1$ to the coset
 $\cU_m$. That is, decompose
\[
\fD_1=T_{\bm{A}}\cU_m\oplus \cV_{\bm{A}},\quad \fD_1^*=\fD_0=
T_{\bm{A}}^*\cU_m\oplus\cV^*_{\bm{A}},
\]
where
\begin{align*}\label{}
&\cV_{\bm{A}}=(\cD^-_1)_{\ge 2m+1}\times ((\cD^+_1)_{<0}/T_\rho\cM),  \\
& \cV^*_{\bm{A}}=(\cD^-_0)_{<-2m-1}\times(T_\rho^*)^\bot \cM, \quad
(T_\rho^*)^\bot \cM=\{\hat{Y}\in(\cD^+_0)_+\mid \hat{Y}(\rho)=0\},
\end{align*}
and then check that the map
\begin{equation*}\label{}
\cP_{\bm{A}}=\left(
              \begin{array}{cc}
                \cP_{\bm{A}}^{\cU\cU} & \cP_{\bm{A}}^{\cU\cV} \\
                \cP_{\bm{A}}^{\cV\cU} & \cP_{\bm{A}}^{\cV\cV} \\
              \end{array}
            \right): T_{\bm{A}}^*\cU_m\oplus \cV^*_{\bm{A}}\to
            T_{\bm{A}}\cU_m\oplus\cV_{\bm{A}}
\end{equation*}
defined in \eqref{PAX} is diagonal.  The lemma is proved.
\end{prf}

The third step is to introduce a shift transformation on the coset
$\cU_m$ as
\begin{equation}
\mathscr{S}: (A,\hat{A})\mapsto (A+s\,D,\hat{A}+s\, D),
\end{equation}
where $s$ is a parameter. The push-forward of the Poisson tensor
$\cP$ in Lemma~\ref{lem-Poicoset} reads
\[
\mathscr{S}_*\cP=\cP_2-s\,\cP_1+s^2\,\cP_0.
\]
It is straightforward to calculate $\cP_1$ and $\cP_0$ (it vanishes
indeed), which leads to the following lemma.

\begin{lem} \label{thm-poi12}
On the coset $\cU_m$ there exist two compatible Poisson tensors:
\begin{align}\label{Poi1}
\cP_1(X,\hat{X})=& \big(-(D X+D\hat{X})_-A-(A X+\hat{A}\hat{X})_-D
\nn\\ &\quad
+ A(X D+\hat{X}D)_-+D(X A+\hat{X}\hat{A})_-, \nn\\
&\quad (D X+D \hat{X})_+\hat{A}+(A X+\hat{A}\hat{X})_+D \nn\\ &\quad
- \hat{A}(X D+\hat{X}D)_+ -D(X A+\hat{X}\hat{A})_+\big), \\
\cP_2(X,\hat{X})=& \big(-(A X+\hat{A}\hat{X})_-A+A(X
A+\hat{X}\hat{A})_-, \nn\\&\quad (A
X+\hat{A}\hat{X})_+\hat{A}-\hat{A}(X A+\hat{X}\hat{A})_+\big)
\label{Poi2}
\end{align}
with $(X,\hat{X})\in T^*_{\bm{A}}\cU_m$ at any point
$\bm{A}=(A,\hat{A})\in\cU_m$.
\end{lem}

Let $\{~,\,\}_{1,2}^m$ denote the Poisson brackets on $\cU_m$ given
by the tensors $\cP_{1,2}$ respectively. Finally, we arrive at the
main theorem of this section.
\begin{thm}\label{thm-Poi2BKP}
For any positive integer $m$, the two-component BKP hierarchy
\eqref{PPht}-- \eqref{PPhth} can be represented in a bi-Hamiltonian
form as
\begin{align}\label{Ham2BKP}
&\frac{\pd F}{\pd t_k}=\{F, H_{k+2m}\}^m_1=\{F, H_{k}\}^m_2, \\
&\frac{\pd F}{\pd \hat{t}_k}=\{F, \hat{H}_{k+2}\}^m_1=\{F,
\hat{H}_{k}\}^m_2 \label{Ham2BKP2}
\end{align}
with $k\in\Zop$ and Hamiltonians
\begin{align}\label{HHhat}
H_{k}=\frac{2\,m}{k}\la P^k\ra, \quad \hat{H}_{k}=\frac{2}{k}\la
\hat{P}^k\ra.
\end{align}
\end{thm}
\begin{prf}
The proof is similar to that of Theorem~5.4 in \cite{WX}, so we only
sketch the main steps. The gradient of the Hamiltonians are
\begin{equation}\label{}
\frac{\dt H_k}{\dt\bm{A}}=(P^{k-2m}D^{-1},0), \quad \frac{\dt
\hat{H}_k}{\dt\bm{A}}=(0, \hat{P}^{k-2}D^{-1})
\end{equation}
up to a kernel part of the form $(Z,\hat{Z})\in(\cD_0^-)_{\le
-2m-2}\times(\cD_0^+)_{\ge0}$ such that $\hat{Z}_+(\rho)=0$. Such a
kernel part does not change the following Hamiltonian equations
\begin{align}\label{}
&\frac{\pd\bm{A}}{\pd t_k}=\cP_1\left(\frac{\dt
H_{k+2m}}{\dt\bm{A}}\right)=\cP_2\left(\frac{\dt
H_{k}}{\dt\bm{A}}\right),
\\
&\frac{\pd\bm{A}}{\pd\hat{t}_k}=\cP_1\left(\frac{\dt
\hat{H}_{k+2}}{\dt\bm{A}}\right)=\cP_2\left(\frac{\dt
\hat{H}_{k}}{\dt\bm{A}}\right).
\end{align}
It is a straightforward calculation to convert these Hamiltonian
flows to the Lax equations \eqref{At}--\eqref{Ath}. Therefore the
theorem is proved.
\end{prf}

Observe the difference between the bi-Hamiltonian structures
\eqref{Ham2BKP}--\eqref{Ham2BKP2} and the one given in \cite{WX}.
Moreover, the densities of Hamiltonian functionals in \eqref{HHhat},
in contrast to those of $H_{k}=\frac{2}{k}\la P^k\ra$ and
$\hat{H}_{k}=-\frac{2}{k}\la \hat{P}^k\ra$ in (5.19) of \cite{WX},
are tau-symmetry, that is, the following $1$-form is closed:
\[
\om=\sum_{k\in\Zop}(\res\,P^k\,\od
t_k+\res\,\hat{P}^k\,\od\hat{t}_k).
\]
Given an arbitrary solution to the two-component BKP hierarchy,
there locally exists a tau function $\tau=\tau(\bm{t},\hat{\bm{t}})$
such that
\begin{align}\label{tau}
\om=\od(2\,\pd_x\log\tau).
\end{align}
Here $\bm{t}=(t_1,t_3,\dots)$ and
$\hat{\bm{t}}=(\hat{t}_1,\hat{t}_3,\dots)$ with $t_1=x$. In
\cite{LWZ} it was showed that the hierarchy
\eqref{PPht}--\eqref{PPhth} is equivalent to the bilinear equation
of the two-component BKP hierarchy \cite{DJKM-KPtype}:
\begin{align} \label{bltau}
&\res_z z^{-1}X(\bm{t};z)\tau(\bm{t},\hat{\bm{t}})
X(\bm{t}';-z)\tau(\bm{t}',\hat{\bm{t}}')\nn\\
=&\res_z z^{-1}X(\hat{\bm{t}};z)\tau(\bm{t},\hat{\bm{t}})
X(\hat{\bm{t}}';-z)\tau(\bm{t}',\hat{\bm{t}}'),
\end{align}
where $X$ is a vertex operator defined by
\[X(\bm{t};z)=\exp\left(\sum_{k\in\Zop} t_k z^k\right)
\exp\left(-\sum_{k\in\Zop}\frac{2}{k z^k}\frac{\pd}{\pd t_k}\right)
\]
and $\res_z\sum f_i z^i=f_{-1}$ for formal series in $z$.

\subsection{Reductions of bi-Hamiltonian structures}

We now study the reduction property of the bi-Hamiltonian structures
given in Theorem~\ref{thm-Poi2BKP}.

First, suppose the pseudo-differential operator $\bm{A}=(A,\hat{A})$
in \eqref{eqbK} satisfies
\begin{equation}\label{red}
A=\hat{A}, \quad \hbox{ i.e., }  \quad P^{2 m}=\hat{P}^2.
\end{equation}
This constraint reduces the Lax equations \eqref{At}--\eqref{Ath} to
\begin{equation}\label{}
\frac{\pd L}{\pd t_k}=D[(P^k)_+,D^{-1}L], \quad \frac{\pd L}{\pd
\hat{t}_k}=D[-(\hat{P}^k)_-,D^{-1}L],
\end{equation}
where $L=A=\hat{A}$ and $k\in\Zop$. This system of Lax equations is
equivalent to the Drinfeld-Sokolov hierarchy associated to the
affine Lie algebra $D_{m+1}^{(1)}$ with the zeroth vertex $c_0$ of
its Dynkin diagram marked \cite{DS, LWZ}.

Note that the operator $L=A=\hat{A}$ lies in the coset
\begin{equation}\label{dsd}
\mathcal{W}_m=D^{2m+1}+\left\{\sum_{i=1}^{m}\left( w_i D^{2(m-i)+1}+
D^{2(m-i)+1} w_i \right)+\rho D^{-1}\rho\in\cD^-\cap\cD^+ \right\}.
\end{equation}
On this coset we let $F_X( L)$ denote the functional that has
gradient $X$ with respect to $L$.
\begin{prp}\label{thm-dham}
The Poisson brackets in Theorem~\ref{thm-Poi2BKP} under the
constraint \eqref{red} are restricted to $\mathcal{W}_m$ as (the
superscript $m$ is omitted):
\begin{align}\label{Poid1}
&\{F_X( L),F_Y( L)\}_1
\nn \\
&\quad =\big\la\left( L X\right)_-\left(D Y\right)_+ + \left(D
X\right)_-\left( L Y\right)_+ - \left( X L\right)_-\left( Y
D\right)_+-\left( X
D\right)_-\left( Y  L\right)_+\big\ra, \\
&\{F_X( L),F_Y( L)\}_2=\big\la\left( L X\right)_-\left( L Y\right)_+
- \left( X L\right)_-\left( Y  L\right)_+\big\ra. \label{Poid2}
\end{align}
They give the bi-Hamiltonian structure for the Drinfeld-Sokolov
hierarchy \eqref{dsd} of type $(D_{m+1}^{(1)},c_0)$ (cf.
Proposition~8.3 of \cite{DS}, see also \cite{DLZ, LWZ}).
\end{prp}
\begin{prf}
Suppose $\dt F_X( L)/\dt\bm{A}=(W,\hat{W})$ whenever $F_X( L)$ is
viewed as a functional on the coset $\cU_m$, then we have
$X=(W+\hat{W})|_{A=\hat{A}=L}$. Thus the proposition follows from
\eqref{Poi1}--\eqref{Poi2}.
\end{prf}


Second, we let $\hat{P}\to0$, namely, the flows \eqref{PPhth} along
$\hat{\bm{t}}$ vanish, then the two-component BKP hierarchy is
reduced to the (one-component) BKP hierarchy \cite{DKJM-KPBKP}.
Similarly as above, we have
\begin{prp}
If $\hat{P}\to0$, then the formulae
\eqref{Ham2BKP}--\eqref{Ham2BKP2} are reduced to bi-Hamiltonian
representations for the BKP hierarchy given by the following Poisson
brackets:
\begin{align} \label{BKPp1}
&\{F_X(A),F_Y(A)\}_1^m=\big\la X\left( -D Y_-A -( A
Y)_-D + A Y_- D+D(Y A)_-\right)\big\ra, \\
&\{F_X( A),F_Y( A)\}_2^m=\big\la-(A X)_+( A Y)_- +( X A)_+( Y
A)_-\big\ra, \label{BKPp2}
\end{align}
where $A=D P^{2m}$.

Furthermore, by setting $A_-=(D P^{2m})_-=0$  one reduces
\eqref{BKPp1}--\eqref{BKPp2} to the bi-Hamiltonian structure for the
Drinfeld-Sokolov hierarchy of type $(B_m^{(1)}, c_0)$ (see
\cite{DS}):
\begin{align}
&\{F_X(L),F_Y(L)\}_1=\big\la L( Y D X-X D Y)\big\ra, \\
&\{F_X( L),F_Y( L)\}_2=\big\la(L X)_-( L Y)_+ -( X L)_-( Y
L)_+\big\ra,
\end{align}
where $L=A=A_+$.
\end{prp}

\subsection{Other Hamiltonian structures and their reductions}

Besides those in \eqref{Ham2BKP}--\eqref{Ham2BKP2}, the $R$-matrix
\eqref{Rmatrix} can produce more Hamiltonian structures for the
two-component BKP hierarchy.

Given any positive integer $m$, we replace \eqref{eqbK} by
$\bm{A}=(D P^{2m-1},D\hat{P})$. All such operators form a coset
\[
\mathcal{V}_m=(D^{2m},0)+(\cD_0^-)_{<2m}\times(\cD_0^+)_{\ge0}.
\]
With the same method as in the previous subsection, one can restrict
the Poisson bracket \eqref{Poibra} properly to the coset
$\mathcal{V}_m$. Let $\{~,\,\}^m$ denote the restricted bracket,
then we have the following proposition.
\begin{prp}\label{thm-PoiBKP}
For any positive integer $m$, the two-component BKP hierarchy has
the following Hamiltonian representation:
\begin{align}\label{Ham2BKP3}
\frac{\pd F}{\pd t_k}=\{F, H_{k}\}^m, \quad \frac{\pd F}{\pd
\hat{t}_k}=\{F, \hat{H}_{k}\}^m, \quad k\in\Zop,
\end{align}
where
\[
H_{k}=\dfrac{2\,m-1}{k}\la P^k\ra, \quad \hat{H}_{k}=\dfrac{1}{k}\la
\hat{P}^k\ra.
\]
\end{prp}

Assume that the Lax equations \eqref{PPht}--\eqref{PPhth} are
constrained by
\begin{equation}\label{red2}
D P^{2m-1}=D\hat{P}.
\end{equation}
The reduced equations compose the Drinfeld-Sokolov \cite{DS}
hierarchy of type $(A_{2m-1}^{(2)},c_0)$, which possesses a
Hamiltonian structure reduced from \eqref{Ham2BKP2} as, with $L=D
P^{2m-1}=D\hat{P}$,
\begin{equation}\label{a2m}
\{F_X( L),F_Y( L)\}=\big\la(L X)_-( L Y)_+ -( X L)_-( Y L)_+\big\ra.
\end{equation}
\begin{exa}
In the particular case of $m=1$ so that $P=D+D^{-1}\rho$, we have
the reduced hierarchy
\begin{equation}\label{kdv}
\frac{\pd P}{\pd t_k}=[(P^k)_+, P], \quad k\in\Zop.
\end{equation}
The first three equations read:
\begin{align*}
&\rho_{t_1}=\rho_x, \quad \rho_{t_3}=6\rho\rho_x+\rho_{xxx}, \\
&\rho_{t_5}=30\rho^2\rho_x+20\rho_x\rho_{xx}+10\rho\rho_{xxx}+\rho_{xxxxx},
\end{align*}
where the subscripts standing for partial derivatives. In this case,
the Poisson bracket \eqref{a2m} can be written as
\[
\{\rho(x),\rho(y)\}=2\rho(x)\,\dt'(x-y)+\rho'(x)\,\dt(x-y)+\frac1{2}\dt'''(x-y),
\]
which is the ``second'' Hamiltonian structure for the Korteweg-de
Vries (KdV) hierarchy (see, for example, \cite{Dickey}). On the
other hand, the Hamiltonians are
\[
H_{k}=\dfrac{1}{k}\la
P^k\ra=\int\left(\frac{1}{k}\binom{k}{(k-1)/2}\rho^{(k+1)/2}+\hbox{
derivative terms } \right)\od x,
\]
Thus the hierarchy \eqref{kdv} is indeed equivalent to the KdV
hierarchy.
\end{exa}

\subsection{Dispersionless case}

We continue to construct Hamiltonian structures for the
dispersionless two-component BKP hierarchy and study their
reductions.

Consider two algebras $\cH^-=\cA((z^{-1}))$ and $\cH^+=\cA((z))$ of
Laurent series in $z\in S^1$. 
On each $\cH^\pm$ there exist a Lie bracket
\begin{equation}\label{}
[a,b]=\frac{\pd a}{\pd z}\frac{\pd b}{\pd x}-\frac{\pd b}{\pd
z}\frac{\pd a}{\pd x},
\end{equation}
and an invariant inner product
\begin{equation}\label{}
\la a, b\ra=\la a\,b\ra, \quad \la
a\ra=\frac1{2\pi\sqrt{-1}}\oint_{S^1}\oint_{S^1}a(z)\,\od z\,\od x.
\end{equation}

Let
\begin{equation}\label{}
p(z)=z+\sum_{i\le0}u_i\,z^{2i-1}\in\cH^-, \quad
\hat{p}(z)=\sum_{i\ge0}\hat{u}_i\,z^{2i-1}\in\cH^+,
\end{equation}
then the dispersionless two-component BKP hierarchy is defined as
\begin{equation}\label{d2BKP}
\frac{\pd \al(z)}{\pd t_k}=[(p(z)^k)_+,\al(z)], \quad \frac{\pd
\al(z)}{\pd \hat{t}_k}=[-(\hat{p}(z)^k)_-,\al(z)], \quad k\in\Zop,
\end{equation}
where $\al(z)=p(z), \hat{p}(z)$. Here the subscripts ``$\pm$'' stand
for the projections of a series in $z$ to its nonnegative and
negative part respectively.
 Recall that the hierarchy
\eqref{d2BKP} was first written down by Takasaki \cite{Ta} as the
hierarchy underlying the D-type topological Landau-Ginzburg models.

Introduce the coupled Lie algebra $\mathfrak{H}=\cH^-\times\cH^+$,
which is equipped with an inner product
\[
\la(a,\hat{a}),(b,\hat{b})\ra=\la a,b\ra+\la\hat{a},\hat{b}\ra,
\quad (a,\hat{a}),~(b,\hat{b})\in \mathfrak{H}.
\]
In fact $\mathfrak{H}$ is a Poisson algebra, hence one can apply
Theorem~\ref{thm-Li} to construct Poisson brackets between
functionals on it with the $R$-matrix $R$ given in \eqref{Rmatrix}.

Given any positive integers $m$ and $n$, let
\begin{equation}\label{}
\bm{a}(z)=(a(z),\hat{a}(z))=(z\,p(z)^{2m},z\,\hat{p}(z)^{2n}).
\end{equation}
All such series form a coset
\begin{equation}\label{}
U_{m,n}=(z^{2m+1},0)+\set{\left(\sum_{i\le
m}v_i\,z^{2i-1},\sum_{i\ge
1-n}\hat{v}_i\,z^{2i-1}\right)\in\cH^-\times\cH^+}.
\end{equation}
\begin{lem}
On the coset $U_{m,n}$ there are two compatible Poisson brackets
$\set{~,~}^{m,n}_\nu$ ($\nu=1,2$) given by the following tensors:
\begin{align}
&\cP_1(X(z),\hat{X}(z))\nn\\
=&\big(-z([a(z) ,X(z) ]_-+[\hat{a}(z),\hat{X}(z)]_-)-a(z)(\pd_x
X(z)+\pd_x\hat{X}(z))_-
\nn\\ &\quad +[a(z),z(X(z)+\hat{X}(z))_-]+\pd_x(X(z) a(z)+\hat{X}(z)\hat{a}(z))_-, \nn\\
&\quad~ z([a(z) ,X(z) ]_+ +[\hat{a}(z),\hat{X}(z)]_+)+\hat{a}(z)(\pd_x X(z)+\pd_x\hat{X}(z))_+ \nn\\
&\quad-[\hat{a}(z),z(X(z)+\hat{X}(z))_+]-\pd_x(X(z)
a(z)+\hat{X}(z)\hat{a}(z))_+\big),
\label{dP1} \\
&\cP_2(X(z) ,\hat{X}(z) )\nn\\
=&\big(-a(z)([a(z) ,X(z) ]_-+[\hat{a}(z),\hat{X}(z)]_-)+[a(z),(X(z)
a(z)+\hat{X}(z)\hat{a}(z))_-], \nn\\
&\quad~ \hat{a}(z)([a(z) ,X(z) ]_+ +[\hat{a}(z),\hat{X}(z)]_+)
-[\hat{a}(z), (X(z) a(z)+\hat{X}(z)\hat{a}(z))_+]\big). \label{dP2}
\end{align}
where $(X(z),\hat{X}(z) )\in T^*_{\bm{a}} U_{m,n}$ are covectors.
\end{lem}
\begin{prf}
The proof is similar with that of Lemma~\ref{lem-Poicoset}. First,
Theorem~\ref{thm-Li} with $R$ in \eqref{Rmatrix} gives a quadratic
Poisson bracket ($r=2$) on $\mathfrak{H}$. Second, this Poisson
bracket is checked to be restricted to the coset $U_{m,n}$. The last
step is to consider the push-forward of the Poisson tensor induced
by the shift transformation on $U_{m,n}$: $\bm{a}(z)\mapsto
\bm{a}(z)+(s z,s z)$, where $s$ is a parameter. The lemma is proved.
\end{prf}

Thus in the same way as before we arrive at
\begin{thm} \label{thm-hamd2kp}
For any positive integers $m$ and $n$, the dispersionless
two-component BKP hierarchy \eqref{d2BKP} can be written as
\begin{align}\label{Hamd2BKP}
&\frac{\pd F}{\pd t_k}=\{F, H_{k+2m}\}^{m,n}_1=\{F, H_{k}\}^{m,n}_2,
\\
&\frac{\pd F}{\pd \hat{t}_k}=\{F, \hat{H}_{k+2n}\}^{m,n}_1=\{F,
\hat{H}_{k}\}^{m,n}_2 \label{Hamd2BKP2}
\end{align}
with $k\in\Zop$ and
\begin{equation*}\label{}
H_{k}=\frac{2\,m}{k}\la p(z)^k\ra, \quad
\hat{H}_{k}=\frac{2\,n}{k}\la \hat{p}(z)^k\ra.
\end{equation*}
\end{thm}

Now we assume
\begin{equation}\label{pphl}
p(z)^{2m}=\hat{p}(z)^{2n}=l(z)
\end{equation}
with
\[
l(z)=z^{2m}+\sum_{i=1-n}^{m}v_i\,z^{2i-2}.
\]
then the dispersionless two-component BKP hierarchy \eqref{d2BKP} is
reduced to the following
\begin{equation}\label{dmn}
\frac{\pd l(z)}{\pd t_k}=[(p(z)^k)_+,l(z)], \quad \frac{\pd
l(z)}{\pd \hat{t}_k}=[-(\hat{p}(z)^k)_-,l(z)], \quad k\in\Zop.
\end{equation}
With similar notations and method for Proposition~\ref{thm-dham}, we
have
\begin{prp}
Under the constraint \eqref{pphl}, the bi-Hamiltonian structure
\eqref{dP1}--\eqref{dP2} is reduced to
\begin{align}\label{Poidmn1}
&\{F_X(l),F_Y(l)\}_1 =\la X(z),[l(z),Y(z)]_+ -[l(z),Y(z)_+] \ra, \\
&\{F_X(l),F_Y(l)\}_2=\la X(z),l(z)[l(z),Y(z)]_+ -[l(z),(l(z)
Y(z))_+]\ra. \label{Poidmn2}
\end{align}
They give a bi-Hamiltonian structure for the reduced hierarchy
\eqref{dmn}.
\end{prp}

The quantization of the dispersionless hierarchy \eqref{dmn} is the
two-component BKP hierarchy \eqref{PPht}--\eqref{PPhth} constrained
by $P^{2 m}=\hat{P}^{2 n}$. This is called the $(2\,m,
2\,n)$-reduction, which corresponds to the reduction of Lie algebras
from $\mathfrak{go}(2\,\infty)$ to $D_{m+n}^{(1)}$ in the notation
of \cite{DJKM-reduce}.

With a dressing method as in \cite{LWZ}, one can show that the
$(2\,m, 2\,n)$-reduction of the two-component BKP hierarchy is
equivalent to the following bilinear equation of tau function:
\begin{align} \label{bltau}
&\res_z z^{2m j-1}X(\bm{t};z)\tau(\bm{t},\hat{\bm{t}})X(\bm{t}';-z)\tau(\bm{t}',\hat{\bm{t}}')\nn\\
=&\res_z z^{2n
j-1}X(\hat{\bm{t}};z)\tau(\bm{t},\hat{\bm{t}})X(\hat{\bm{t}}';-z)\tau(\bm{t}',\hat{\bm{t}}'),
\quad j\ge0.
\end{align}
Note that the bilinear equation \eqref{bltau} with $j=0$ is the
original form of the two-component BKP hierarchy \cite{DJKM-KPtype},
and the case $j=1$ was written down in \cite{DJKM-reduce} (see
equation (2.25) there).

When $n=1$, the bilinear equation \eqref{bltau} coincides with the
 Drinfeld-Sokolov hierarchy of type
$(D_{m+1}^{(1)},c_0)$, whose bi-Hamiltonian structure
\eqref{Poid1}--\eqref{Poid2} can be reduced from that of the
two-component BKP hierarchy. When $n>1$, however, up to now we only
obtain the bi-Hamiltonian structure \eqref{Poidmn1}--\eqref{Poidmn2}
for the dispersionless Lax equations. The difficulty in studying the
dispersive case is the lack of a clear description for the manifold
composed by operators of the form $(\hat{P}^{2n})_-$ with $\hat{P}$
given in \eqref{PhP}.

\begin{rmk}
In \cite{WX2} we associated each bi-Hamiltonian structure in
Theorem~\ref{thm-hamd2kp} to an infinite-dimensional Frobenius
manifold $\cM_{m,n}$ consisting of Laurent series of the form
$(p(z)^{2m}, \hat{p}(z)^{2n})$. We also showed that $\cM_{m,n}$
contains an $(m+n)$-dimensional Frobenius submanifold $M_{m,n}$ for
the Coxceter group $B_{m+n}$, which is associated with the
bi-Hamiltonian structure \eqref{Poidmn1}--\eqref{Poidmn2}, as well
as the principal hierarchy \eqref{dmn}. From this point of view, the
reductions of such bi-Hamiltonian structures can be interpreted by
Frobenius submanifolds.
\end{rmk}

\section{Hamiltonian structures for the Toda lattice hierarchy}
\label{sec-TL}

Let us apply the $R$-matrix formalism to the Toda lattice hierarchy.

\subsection{Toda lattice hierarchy}

Assume $\cA$ to be the set of discrete functions with compact
support on $\Z$, and $\Ld$ be a shift operator on $\cA$ such that
$\Ld(f(n))=f(n+1)$. Denote
\[
\cE=\set{\sum_{i\in\Z}f_i\,\Ld^i\mid f_i\in\cA}.
\]
For $A=\sum_{i\in\Z}f_i\,\Ld^i\in\cE$ one has the following
notations:
\begin{align}\label{}
& A_{\ge k}=A_{>k-1}=\sum_{i\ge k}f_i\,\Ld^i, \quad A_{<k}=A_{\le
k-1}=\sum_{i<k}f_i\,\Ld^i, \\
& \Res\,A=f_0, \quad
\pair{A}=\sum_{n\in\Z}\Res\,A(n)=\sum_{n\in\Z}f_0(n).
\end{align}
Consider the following two subspaces of $\cE$:
\begin{equation}\label{}
\cE^-=\set{ \sum_{i<\infty} f_i\, \Ld^i \mid f_i\in\cA}, \quad
\cE^+=\set{ \sum_{i>-\infty} f_i\, \Ld^i \mid f_i\in\cA}.
\end{equation}
On each of $\cE^\pm$ one introduces a product defined by
\[
f(m)\Ld^i\cdot g(n)\Ld^j=f(m)g(n+i)\Ld^{i+j},
\]
then they become associative algebras, and they are Lie algebras
with Lie bracket given by the commutator. It can be checked that
$\la [A,B]\ra=0$ for any $A, B$ in $\cE^-$ or $\cE^+$, hence on each
of $\cE^\pm$ there is an invariant inner product
\begin{equation}\label{proEpm}
 \pair{A,B}=\pair{A\,B}=\pair{B\,A}.
\end{equation}

Introduce
\begin{equation} \label{LhL}
L= \Ld+\sum_{i\le0}u_i  \Ld^{i}\in\cE^-, \quad \hat{L}=
\sum_{i\ge-1}\hat{u}_i \Ld^i\in\cE^+
\end{equation}
with unknown functions $u_i$ and $\hat{u}_i$ lying in $\cA$.
\begin{dfn}
The Toda lattice hierarchy is defined as \cite{UT}
\begin{align}\label{LLht}
& \frac{\pd L}{\pd t_k}=[(L^k)_{\ge0}, L], \quad \frac{\pd \hat{L}}{\pd t_k}=[(L^k)_{\ge0}, \hat{L}],  \\
\label{LLhth} & \frac{\pd L}{\pd \hat{t}_k}=[-(\hat{L}^k)_{<0}, L],
\quad \frac{\pd \hat{L}}{\pd \hat{t}_k}=[-(\hat{L}^k)_{<0},
\hat{L}],
\end{align}
where $k$ runs over all positive integers.
\end{dfn}
Recall that the subscripts ``$\ge0$'' and ``$<0$'' mean the
projections induced by the following decompositions of Lie
subalgebras:
\[
\cE^\pm=(\cE^\pm)_{\ge0}\oplus(\cE^\pm)_{<0}.
\]

\subsection{Hamiltonian structures}

The coupled Lie algebra $\fg$ in Section~3 can be realized as
\begin{equation}\label{}
\fE=\cE^-\times\cE^+,
\end{equation}
whose Lie bracket is defined diagonally. On $\fE$ there is an inner
product induced by \eqref{proEpm} in the same way as \eqref{inpD}.

Consider functionals of the form $F(\bm{A})=\sum_{n\in\Z}f(n)$,
where $f$ is a discrete function depending on $\bm{A}\in\fE$. Here
we restrict ourselves to the functionals whose gradient lies in
$\fE$, that is, there exists $\bm{X}\in\fE$ such that $\dt
F(\bm{A})=\la \dt \bm{A}, \bm{X}\ra$; in this case we write $\dt
F(\bm{A})/\dt \bm{A}=\bm{X}$.

The $R$-matrix \eqref{Rmatrix} on the Lie algebra $\fE$ is
\begin{equation}\label{}
R(X,\hat{X})=(X_{\ge0} -X_{<0} -2\hat{X}_{<0},\hat{X}_{\ge0}
-\hat{X}_{<0} +2 X_{\ge0}).
\end{equation}
Its adjoint transformation $R^*$, satisfying $\la R(X,\hat{X}),
(Y,\hat{Y})\ra=\la (X,\hat{X}), R^*(Y,\hat{Y})\ra$,  reads
\begin{align*}\label{}
R^*(X,\hat{X})=-R(X,\hat{X})+2\,R_0(X,\hat{X}),
\end{align*}
where
\[
R_0(X,\hat{X})=(\Res(X+\hat{X}),\Res(X+\hat{X})).
\]
Thus the anti-symmetric part of $R$ is
\begin{equation}\label{RsTD}
R_a(X,\hat{X})=\frac1{2}(R(X,\hat{X})-R^*(X,\hat{X}))=R(X,\hat{X})-R_0(X,\hat{X}).
\end{equation}
\\
{\bf Claim~} The transformation $R_a$ satisfies the modified
Yang-Baxter equation \eqref{YBeq} on $\cE$.
\\
\\
\begin{prf}
Sine $R$ is a solution of the modified Yang-Baxter equation, then
for any $\bm{X}=(X,\hat{X}), \bm{Y}=(Y,\hat{Y})\in\fE$, we have
\begin{align}
&[R_a(\bm{X}),R_a(\bm{Y})]-R_a([R_a(\bm{X}),\bm{Y}]+[\bm{X},R_a(\bm{Y})])+[\bm{X},\bm{Y}]
\nn\\
=&-[R_0(\bm{X}),R(\bm{Y})]-[R(\bm{X}),R_0(\bm{Y})]+
R([R_0(\bm{X}),\bm{Y}]+[\bm{X},R_0(\bm{Y})]) \nn\\
& +R_0([R(\bm{X}),\bm{Y}]+[\bm{X},R(\bm{Y})])
-R_0([R_0(\bm{X}),\bm{Y}]+[\bm{X},R_0(\bm{Y})]). \label{eq-cl}
\end{align}
On the right-hand side, the first three terms cancel by using
\[
[R_0(\bm{X}),R(\bm{Y})]=R([R_0(\bm{X}),\bm{Y}]),
\]
the fourth term is equal to $(f,f)$ with
\begin{align*}
f=&\Res([X_{\ge0} -X_{<0} -2\hat{X}_{<0},Y]+[X,Y_{\ge0} -Y_{<0}
-2\hat{Y}_{<0}]) \\
&+ \Res( [\hat{X}_{\ge0} -\hat{X}_{<0} +2
X_{\ge0},\hat{Y}]+[\hat{X},\hat{Y}_{\ge0}
-\hat{Y}_{<0} +2 Y_{\ge0}]) \\
=& 2\,\Res([X_{\ge0}-\hat{X}_{<0},Y]+[X,-Y_{<0} -\hat{Y}_{<0}] \\
& \quad + [\hat{X}_{\ge0}+X_{\ge0},\hat{Y}]+[\hat{X},-\hat{Y}_{<0} +
Y_{\ge0}])
\\=& 2\,\Res([X, Y_{\le0}]-[\hat{X},Y_{>0}]+[X,-Y_{<0} -\hat{Y}_{<0}] \\
& \quad +[\hat{X}+X,\hat{Y}_{\le0}]+[\hat{X},-\hat{Y}_{<0} +
Y_{\ge0}])\\
=&\Res([X+\hat{X},\Res(Y+\hat{Y})])=0,
\end{align*}
and clearly the last term vanishes. Thus the claim is verified.
\end{prf}

According to Theorem~\ref{thm-3br}, we have the following lemma.
\begin{lem}\label{thm-PE}
For arbitrary functionals $F$ and $H$ on $\fE$, there exist three
compatible Poisson brackets:
\begin{equation}\label{}
\{F, H\}_{\nu}(\bm{A})=\left\la\frac{\dt F}{\dt\bm{A}},
\cP_\nu\left(\frac{\dt H}{\dt\bm{A}}\right)\right\ra, \quad
\nu=1,2,3,
\end{equation}
where $\bm{A}\in\fE$ and the Poisson tensors $\cP_\nu: T\fD^*\to
T\fD$ read
\begin{align}
\cP_1(X,\hat{X})=&\big([-X_{<0}-\hat{X}_{<0},A]+[X,A]_{\le0}+[\hat{X},\hat{A}]_{\le0}, \nn \\
&\quad
[X_{\ge0}+\hat{X}_{\ge0},\hat{A}]-[X,A]_{>0}-[\hat{X},\hat{A}]_{>0}\big),
\label{P1E}\\
\cP_2(X,\hat{X})=& \frac1{2}\big([-(A X+X A)_{<0}-(\hat{A} \hat{X}+\hat{X} \hat{A})_{<0},A] \nn \\
&\quad +A([X,A]_{\le0}+[\hat{X},\hat{A}]_{\le0})+ ([X,A]_{\le0}+[\hat{X},\hat{A}]_{\le0})A, \nn \\
&\quad [(A X+X A)_{\ge0}+(\hat{A} \hat{X}+\hat{X} \hat{A})_{\ge0},\hat{A}] \nn \\
&\quad-\hat{A}([X,A]_{>0}+[\hat{X},\hat{A}]_{>0})-
([X,A]_{>0}+[\hat{X},\hat{A}]_{>0})\hat{A} \big).
\label{P2E} \\
\cP_3(X,\hat{X})=& \big([-(A X A+\hat{A} \hat{X} \hat{A})_{<0},A]+A([X,A]_{\le0}+[\hat{X},\hat{A}]_{\le0}) A, \nn \\
&\quad [(A X A+\hat{A} \hat{X} \hat{A})_{\ge0},\hat{A}]
-\hat{A}([X,A]_{>0}+[\hat{X},\hat{A}]_{>0})\hat{A} \big).
\label{P3E}
\end{align}
\end{lem}

We want to reduce these Poisson structures to some appropriate
subsets of $\fE$ on which the Lax equations
\eqref{LLht}--\eqref{LLhth} are defined.

Given two arbitrary positive integers $N$ and $M$, with $L$ and
$\hat{L}$ introduced in \eqref{LhL} we let
\begin{equation}\label{ALLh}
\bm{A}=(A,\hat{A})=(L^N,\hat{L}^M).
\end{equation}
All such operators form a coset of $\fE$:
\[
\cU_{N,M}=(\Ld^N,0)+(\cE^-)_{<N}\times(\cE^+)_{\ge-M}.
\]
On this coset, the tangent bundle $T\cU_{N,M}$  and the cogtangent
bundle $T^*\cU_{N,M}$ have their fibers respectively
\begin{align*}\label{}
T_{\bm{A}}\cU_{N,M}=(\cE^-)_{<N}\times(\cE^+)_{\ge-M}, \quad
T_{\bm{A}}^*\cU_{N,M}=(\cE^-)_{>-N}\times(\cE^+)_{\le M}.
\end{align*}

\begin{lem}
On the coset $\cU_{N,M}$ there are two compatible Poisson structures
\begin{equation}\label{}
\cP^{\mathrm{red}}_{\nu}:T^*\cU_{N,M}\to T\cU_{N,M}, \quad \nu=1,2
\end{equation}
defined as
\begin{align}\label{poi1T}
&\cP^{\mathrm{red}}_1(X,\hat{X})=\cP_1(X,\hat{X}), \\
&\cP^{\mathrm{red}}_2(X,\hat{X})=\cP_2(X,\hat{X})-([f,A],[f,\hat{A}]).
\label{poi2T}
\end{align}
where $(X,\hat{X})\in T_{\bm{A}}^*\cU_{N,M}$ and
\begin{equation*}\label{}
f=\frac1{2}(1+\Ld^N)(1-\Ld^{N})^{-1}(\Res([X,A]+[\hat{X},\hat{A}))
\end{equation*}
with $(1-\Ld^{N})^{-1}=1+\Ld^{N}+\Ld^{2 N}+\cdots$.
\end{lem}
\begin{prf}
We need to perform a Dirac reduction for the Poisson structures
$\cP_\nu$ in Lemma~\ref{thm-PE} from $\fE$ to the coset $\cU_{N,M}$.
Let us sketch the main steps (cf. \cite{Carlet}). First, at any
point $\bm{A}\in\cU_{N,M}$ we have the decompositions of subspaces
\[
\cE=T_{\bm{A}}\cU_{N,M}\oplus\cV_{N,M}=T_{\bm{A}}^*\cU_{N,M}\oplus\cV_{N,M}^*,
\]
where
\[
\cV_{N,M}=(\cE^-)_{\ge N}\times(\cE^+) _{<-M}, \quad
\cV_{N,M}^*=(\cE^-)_{\le-N}\times(\cE^+)_{>M}.
\]
Then, the Poisson tensors
\begin{equation*}\label{}
\cP_\nu=\left(
              \begin{array}{cc}
                \cP_\nu^{\cU\cU} & \cP_\nu^{\cU\cV} \\
                \cP_\nu^{\cV\cU} & \cP_\nu^{\cV\cV} \\
              \end{array}
            \right): T_{\bm{A}}^*\cU_{N,M}\oplus\cV_{N,M}^*\to
            T_{\bm{A}}\cU_{N,M}\oplus\cV_{N,M}
\end{equation*}
are reduced to $\cU_{N,M}$ as
\[
\cP_\nu^{\mathrm{red}}=\cP_\nu^{\cU\cU}-
\cP_\nu^{\cU\cV}\circ\left(\cP_\nu^{\cV\cV}\right)^{-1}\circ
\cP_\nu^{\cV\cU}.
\]
After a long but straightforward calculation, we conclude that, the
first tensor $\cP_1$ can be restricted to the coset directly, for
$\cP_2$ one needs a correction term given in \eqref{poi2T} (the
reduction of $\cP_3$ is not clear except for $N=M=1$, see below).
The lemma is proved.
\end{prf}

Let $\{~,\,\}^{N,M}_\nu$ be the Poisson brackets on  the coset
$\cU_{N,M}$ given by the tensors $\cP^{\mathrm{red}}_\nu$ in the
above lemma. The following result can be verified with the same
method as for Theorem~\ref{thm-Poi2BKP}.
\begin{thm}\label{thm-Poi2Toda}
Given any positive integers $N$ and $M$, the  Toda lattice hierarchy
\eqref{LLht}--\eqref{LLhth} has the following bi-Hamiltonian
representation: for $k>0$,
\begin{align}\label{HamT}
&\frac{\pd F}{\pd t_k}=\{F, H_{k+N}\}^{N,M}_1=\{F, H_{k}\}^{N,M}_2,
\\
&\frac{\pd F}{\pd \hat{t}_k}=\{F, \hat{H}_{k+M}\}^{N,M}_1=\{F,
\hat{H}_{k}\}^{N,M}_2
\end{align}
with arbitrary functional $F$ and Hamiltonians
\begin{equation}\label{HHhatT}
H_{k}=\frac{N}{k}\la L^k\ra, \quad \hat{H}_{k}=\frac{M}{k}\la
\hat{L}^k\ra.
\end{equation}
\end{thm}

\begin{rmk}
The densities of the Hamiltonian functionals \eqref{HHhatT} are
tau-symmetric, hence they define a tau function of the Toda Lattice
hierarchy in a similar way as for the two-component BKP hierarchy.
\end{rmk}

When $N=M=1$, on the coset $\cU_{1,1}$ there exists another Poisson
structure $\cP^{\mathrm{red}}_3$ that is compatible with
$\cP^{\mathrm{red}}_1$ and $\cP^{\mathrm{red}}_2$. More precisely,
\begin{equation}\label{poi3T}
\cP^{\mathrm{red}}_3(X,\hat{X})=\cP_3(X,\hat{X})-([Z,A],[Z,\hat{A}]),\quad
(X,\hat{X})\in T_{\bm{A}}^*\cU_{N,M}
\end{equation}
where $Z=(A(g\,\Ld^{-1}+h\,\Ld^{-2})A)_{\ge0}$ with functions $g$
and $h$ determined by
\begin{align*}\label{}
&(1-\Ld)(g)=\Res([X,A]+[\hat{X},\hat{A}]), \\
&(1-\Ld)(h)-g(1-\Ld^{-1})(\Res\,A)=\Res([X,A]\Ld+[\hat{X},\hat{A}]\Ld).
\end{align*}
In this case, the derivatives $\pd/\pd t_k$ and $\pd/\pd\hat{t}_k$
with $k\ge2$ of the Toda lattice hierarchy can also be represented
into Hamiltonian flows of $\cP^{\mathrm{red}}_3$.

In the case $N=M=1$, Carlet \cite{Carlet} derived three Hamiltonian
structures for the Toda lattice hierarchy. Note the difference
between them and $\cP_\nu^{\mathrm{red}}$ in \eqref{poi1T},
\eqref{poi2T} and \eqref{poi3T}.

\subsection{Hamiltonian structures for the extended bigraded Toda hierarchies}

Suppose the  Toda lattice hierarchy is constrained by
\begin{equation}\label{LNLhM}
L^N=\hat{L}^M=\cL,
\end{equation}
where $\cL$ has the form
\begin{equation}\label{LMN}
\cL=\Ld^N+v_{N-1}\Ld^{N-1}+v_{N-2}\Ld^{N-2}+\cdots v_{-M}\Ld^{-M}.
\end{equation}
In the same way as before, under the constraint \eqref{LNLhM} the
Poisson brackets in Theorem~\ref{thm-Poi2Toda} are reduced to:
\begin{align}\label{PoiT1}
\{F_X(\cL),F_Y(\cL)\}_1=&\pair{X,[Y_{\ge0},\cL]_{\le0}-[Y_{<0},\cL]_{>0}}, \\
\{F_X(\cL),F_Y(\cL)\}_2=&\pair{X,[-(\cL\,Y+Y\cL)_{<0},\cL]+\frac1{2}
\cL[Y,\cL]_{\le0}+\frac1{2}[Y,\cL]_{\le0}\cL}\nn\\
&-\pair{X,\frac1{2}[(1+\Ld^N)(1-\Ld^N)^{-1}(\Res\,[Y,\cL])}.
\label{PoiT2}
\end{align}
These formulae have the same expression as the bi-Hamiltonian
structure for the extended $(M,N)$-bigraded Toda hierarchy
\cite{CDZ, Ca06}.

We recall briefly the construction of the extended bigraded Toda
hierarchy. First, a continuation needs to be performed, that is,
$\cA$ must be replaced by an algebra of analytic functions of a
spacial variable $x$, meanwhile the shift operator $\Ld$ replaced by
$e^{\ep D}$ with $D=\od/\od x$ and a small constant $\ep$. Second,
in order to obtain a complete integrable hierarchy starting from
$\cL$ in \eqref{LMN}, one needs not only the roots $L=\cL^{1/N}$ and
$\hat{L}=\cL^{1/M}$ of the form \eqref{LhL}, but also a logarithm
operator $\mathrm{Log}\,\cL$ defined in a dressing way, see
\cite{CDZ, Ca06} for details. Thus up to a scalar transformation of
the time variables, the extended $(N,M)$-bigraded Toda hierarchy is
composed of the Hamiltonian flows given by the Poisson brackets
\eqref{PoiT1}--\eqref{PoiT2} together with Hamiltonians
\eqref{HHhatT} and
\begin{equation}\label{}
H_{k}^L=\frac{2}{(k-1)!}\left\la
\cL^{k-1}\left(\mathrm{Log}\,\cL-\frac1{2}
                   \left(\dfrac1{M}+\dfrac1{N}\right)c_{k-1}\right)\right\ra,
                   \quad
                   k\ge1,
\end{equation}
where $c_0=0$ and $c_{k}=1+1/2+\cdots+1/k$ for $k\ge1$.

One can also consider Hamiltonian structures of the dispersionless
Toda hierarchy. However, in contrast to the previous section,
nothing is obtained beyond the dispersionless limit of the above
result. Note that the dispersionless limit of $\cP^{\mathrm{red}}_2$
for $N=M=1$ was written down in \cite{CDM} (see Proposition~3.3
there) for the purpose of constructing an infinite-dimensional
Frobenius manifold. It is interesting to generalize their
construction to the case of arbitrary positive integers $N$ and $M$.

\section{Conclusion}

On a Lie algebra $\fg$ of the form \eqref{g}, we have obtained a
list of $R$-matrices that satisfy the modified Yang-Baxter equation.
Among these $R$-matrices, the one \eqref{Rmatrix} is selected to
construct Hamiltonian structures for Lax equations defined on $\fg$,
and these Hamiltonian structures have natural but important
reductions. In the examples of the two-component BKP and the Toda
lattice hierarchies, the bi-Hamiltonian structures and their
reductions are closely related to the theory of Frobenius manifold.
In follow-up work we will try to apply the $R$-matrix formalism to
study other hierarchies like multicomponent Toda lattice hierarchies
\cite{UT} and their generalizations.

\vskip 0.4truecm \noindent{\bf Acknowledgment.} The author is
grateful to Boris Dubrovin, Si-Qi Liu and Youjin Zhang for helpful
discussions and comments. The research leading the this work has
received specific funding under the ``Young SISSA Scientists'
Research Projects'' scheme 2012--2013, promoted by the International
School for Advanced Studies (SISSA), Trieste, Italy.

\end{document}